\documentclass[12pt]{iopart}
\usepackage{graphicx,iopams}

\begin{document}

\title{Fulde-Ferrell Pairing Instability in Spin-Orbit Coupled Fermi Gas}

\date{\today}

\author{Lin Dong$^{1}$, Lei Jiang$^{2}$, and Han Pu$^{1}$}

\address{$^{1}$Department of Physics and Astronomy, and Rice Quantum Institute,
Rice University, Houston, TX 77251, USA}
\address{$^{2}$Joint Quantum Institute, University of Maryland and National
Institute of Standards and Technology, Gaithersburg, Maryland 20899,
USA}
%\address{$^{3}$Key Laboratory of Quantum Information, University of Science and Technology of China, CAS, Hefei, Anhui 230026, People's Republic of China}
%
%\affiliation{$^{1}$Department of Physics and Astronomy, and Rice Quantum Institute,
%Rice University, Houston, TX 77251, USA \\
% $^{2}$Joint Quantum Institute, University of Maryland and National
%Institute of Standards and Technology, Gaithersburg, Maryland 20899,
%USA \\
%$^{3}$Key Laboratory of Quantum Information, University of Science and Technology of China, CAS, Hefei, Anhui 230026, People's Republic of China}

\begin{abstract}
We consider finite-momentum pairing of a superfluid ultracold Fermi gas subjected to spin-orbit coupling and an effective Zeeman field. Based on our two-body and mean-field many-body calculations, we show that the Fulde-Ferrell type superfluid dominates in both the zero and the finite temperature phase diagram. We examine the origin and properties of this novel phase systematically. 
\end{abstract}


\submitto{\NJP}


\maketitle

\section{Introduction}

The nature and the microscopic origin of fermionic pairing was first elucidated in
the pioneering work by Bardeen, Cooper, and Schrieffer, widely known
as the BCS theory \cite{BCS}. The attractive pairwise
interaction between electrons with opposite spin, albeit extremely weak, can give rise to an instability
in normal electron gas towards the formation of zero-momentum Cooper pairs near
Fermi surface, and because of pair condensation, ordering of conduction
electrons emerge naturally. When subjected to an external Zeeman field, the population balance between electrons with different spins may be broken. As a consequence,
not all electrons can find a partner to pair up with. If spin-population imbalance is large enough, the pairing of fermions has to occur at finite center-of-mass momentum with deformed Fermi surface state \cite{sed05}. This exotic
possibility of inhomogeneous superfluid was first predicted
by Fulde and Ferrell (FF) \cite{FF}, and by Larkin and
Ovchinnikov (LO) \cite{LO} a little later. FF refers
to an order parameter with plane-wave form $\Delta({\bf r})=\Delta_{0}e^{i{\bf q}\cdot{\bf r}}$,
which spontaneously breaks time-reversal symmetry; while LO considers
the superfluid with a standing-wave order parameter $\Delta({\bf r})=\Delta_{0}\cos({\bf q}\cdot{\bf r})$,
which explicitly breaks translational symmetry. Both phases have puzzled the solid-state
community for decades in terms of unambiguous experimental evidence
to prove their existence. Moreover, the FFLO state is also of interest in quantum chromodynamics at low temperature and high  density, where the property of asymptotic freedom may favor color superconductivity \cite{FFLOrmpQCD}.

In recent years, due to their exquisite controllability, ultracold
atoms have emerged as an ideal platform to simulate many-body Hamiltonians.
Adjustable interaction and high degrees of control over spin-populations
have enabled one with the feasibility of exploring the long sought FFLO
phase. Tremendous theoretical and experimental efforts have been
put into optimizing the best detectable parameter regime of this phase. The most
promising route is now believed to probe the one-dimensional (1D) spin-imbalanced
Fermi gas \cite{kun, feiguin, 1d, erich,Orso, Hui2007, Ueda08, honglu,exp,exp1, guan}, where indirect evidence of FFLO phase has been found in a recent experiment \cite{1DRICE}. However, for 3D Fermi gas, the FFLO phase is not favored \cite{3DMIT, 3DRICE, parish07,note}. 

Over the past few years, another milestone achievement in cold atom research is the realization of artificial spin-orbit (SO) coupling, first in bosonic systems \cite{lin} and later in fermioinic ones \cite{Zhang, mit}. By tailoring the laser fields that generate the SO coupling, various coupling schemes can be realized in principle. It has been realized very recently that, in a Fermi gas, the interplay between the SO coupling and an effective Zeeman field may lead to distortion of single-particle dispersion as well as the Fermi surface, in such a way that finite-momentum dimer state and/or Cooper pairs will be favored \cite{2body,2body1,Chuanweiab,yiwei,
HuiNist}. In this work, we provide a unified treatment of both two-body and the many-body physics for a Fermi gas subjected to an isotropic three-dimensional SO coupling (3DSOC) and an effective Zeeman field. The generation of such 3DSOC has been recently
proposed by optically dressing four internal atomic states with a
tetrahedral geometry \cite{3DSOC, zhou}. This version
of the SO coupling is less explored and unfamiliar to the condensed matter
community, where 2D Rashba and Dresselhaus SO couplings are studied
extensively. One important advantage of 3DSOC over lower-dimensional SO interaction is that it provides the greatest enhancement of fermionic pairing \cite{2body}. Furthermore, due to its isotropic nature, mathematical simplicity is ensured.

The main findings of our work are: i) Under arbitrarily weak Zeeman field, zero-momentum dimer state and conventional BCS superfluid phase are no longer stable. ii) For many-body system, FF state is inherently robust, and ultimately connects to normal phase in a smooth manner as Zeeman field strength is increased, cf.~Fig.~\ref{fig3}. 
%We found consistent results that FF pairing is energetically more favorable in major part of the phase
%diagram, cf. Fig.~\ref{fig3},. More specifically,  FF superfluid is not only favorable
%at sufficiently large Zeeman field but also survives at arbitrarily
%small magnitude, which we call Fulde-Ferrell \emph{pairing
%instability}. 
Moreover, this type of exotic superfluid has a
different origin in comparison with the previously studied FFLO state.
%, which concerns about imbalanced Fermi gas. 
In the absence of the SO coupling,
individual particle number with different spins is conserved, hence
the imbalance induced finite-momentum pairing has parity symmetry
between ${\bf q}$ and $-{\bf q}$, which should be called LO phase
by definition; on the other hand, in the presence of the SO coupling, Zeeman
field breaks time-reversal symmetry explicitly and cause
the single-particle dispersion to be asymmetric, which underlines the idea of finite-momentum dimer bound state \cite{2body} and the FF pairing instability. The center-of-mass momentum of the Cooper pair can be as large as the Fermi momentum.
This result should be very encouraging for future experimental exploration.
 
%At the time of preparing this manuscript, we become
%aware of several relevant works \cite{Chuanweiab,yiwei,
%HuiNist} posted on arXiv.  For those systems, there could be complications arising
%from the competition between out-of-plane and in-plane Zeeman fields.
%However, we focus on the simplest case where isotropic three-dimensional SO coupling
%(3DSOC) is present and Zeeman field is added. The 3DSOC has been recently
%proposed by optically dressing four internal atomic states with a
%tetrahedral geometry \cite{3DSOC, zhou}. This version
%of SO coupling is less explored and unfamiliar to condensed matter
%community, where 2D Rashba and/or Dresselhaus SO couplings are considered
%largely. However, there have been more and more efforts that is of particular importance to understand 3D topological
%insulators \cite{3Dtop},  and Weyl semi-metals \cite{weyla,weylb}. 

The rest of the paper is organized as follows. In Sec.~\ref{model}, we formulate the physical model and introduce the functional path integral technique. We apply this general formalism to the system with 3DSOC and discuss our results on two-body physics in Sec.~\ref{resultsa} and on many-body physics in Sec.~\ref{resultsb}. And finally we conclude in Sec.~\ref{Conclusion}.

\section{Physical model and general formalism}\label{model}

In this section, we first present the model Hamiltonian under study,
and then introduce the widely used functional path integral approach.
Using this approach, we can discuss both the two-body physics and the many-body physics at both
zero and finite temperatures in a unified way.

\subsection{Model Hamiltonian}\label{H0}

We start by formulating the Hamiltonian for a non-interacting homogeneous spin-1/2
Fermi gas in 3D: 
\begin{equation}
\mathcal{H}_{0}=\int d{\bf r}\, \psi^{\dagger}({\bf r})\{\xi_{{\bf k}}+\sum_{i=x,y,z}\left(v_{i}k_{i}+\Lambda_{i}\right)\sigma_{i}\}\psi({\bf r}) \label{eq:modelSingleH0}
\end{equation}
where $\xi_{{\bf k}}=\hbar^{2}\hat{{\bf k}}^{2}/(2m)-\mu$ and 
$\psi=[\psi_{\uparrow}({\bf r}),\psi_{\downarrow}({\bf r})]^{T}$ is the
fermionic annihilation field operator.
We have defined SO coupling strength vector ${\bf v}=(v_{x},v_{y},v_{z})$,
and the Zeeman field vector ${\bf \Lambda}=(\Lambda_{x},\Lambda_{y},\Lambda_{z})$.
${\boldsymbol \sigma}=(\sigma_{x},\sigma_{y},\sigma_{z})$ are Pauli matrices
acting on the atomic (pseudo-)spin degrees of freedom. This description
is a general model valid for various SO coupling schemes. The single-particle spectrum is given by $E^\gamma({\bf k})=\frac{\hbar^2 {\bf k}^2}{2m}+\gamma\sqrt{\sum_i(v_{i}k_{i}+\Lambda_{i})^2}$ with $\gamma=\pm1$ denoting the two helicity branches. The experimentally realized \cite{lin, Zhang, mit} equal
weight Rashba-Dresselhaus SO coupling takes the form of Eq.~(\ref{eq:modelSingleH0})
with ${\bf v}=(0,0,\hbar^{2}k_{r}/m)$ and ${\bf \Lambda}=(\Omega/2,0,\delta)$  where $k_r$ is the laser recoil momentum, $\Omega$ the Raman laser coupling strength, and $\delta$ the two-photon detuning. The
Rashba SO coupling \cite{Jacob2007, RMP} can be recognized
with ${\bf v}=(v_{x},\, v_{y},\,0)$ and $v_{x}=v_{y}=\hbar^{2}k_{r}/m$. In our work, we will focus on the 
3DSOC \cite{3DSOC, zhou} with $v_{x}=v_{y}=v_{z}=v$. For this case,
due to the isotropic nature of the SO coupling term, the direction of
the Zeeman field is irrelevant and we shall choose it to be along the $z$-axis, and hence  ${\bf \Lambda}=(0,\,0,\, h)$. 

\begin{figure}
\begin{center}
\includegraphics[width=0.88\textwidth]{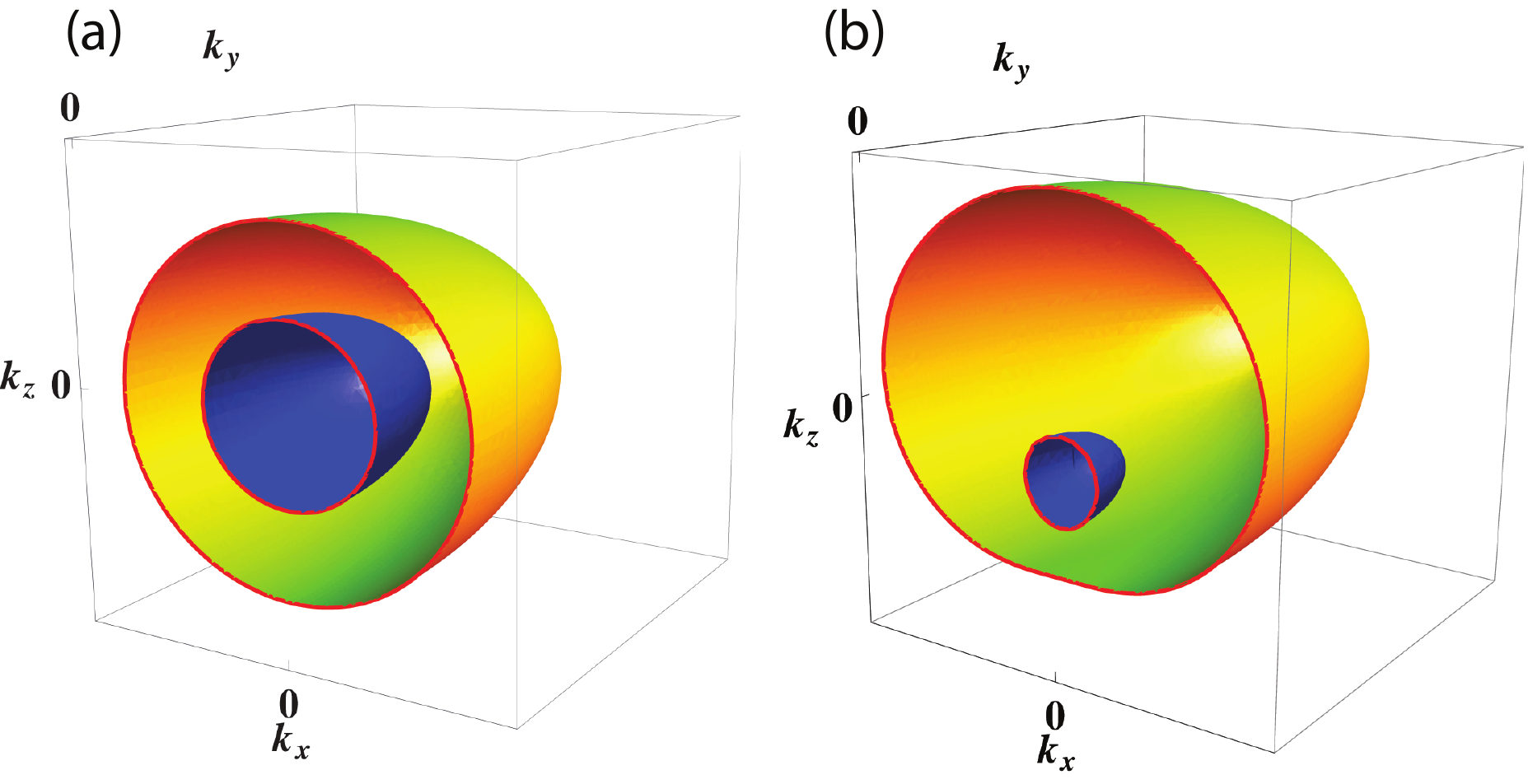}
\end{center}
\caption{(a) Fermi surfaces (a cut in the $k_y=0$ plane) in the absence of the Zeeman field. The two concentric Fermi surfaces are spherically symmetric. The inner blue sphere represents the Fermi surface of the $+$ helicity branch, while the outer yellow sphere of the $-$ helicity branch. (b) Fermi surfaces in the presence of the Zeeman field along the $z$-axis: both Fermi surfaces are deformed in such a way that the cylindrical symmetry about the $k_z$-axis is still preserved, but the reflection symmetry about the $k_z=0$ plane is broken. }\label{FS}
\end{figure}

It is important to note that the Zeeman field does induce an asymmetry in the single-particle dispersion relation. To illustrate this, we consider a filled Fermi sea with simple topology (cf. \cite{shenoyFS}) at zero temperature. In Fig.~\ref{FS}, we plot Fermi surface without and with Zeeman field. In the absence of the Zeeman field, the Fermi surfaces for both helicity branches are represented by spheres centered at zero momentum, as shown in Fig.~\ref{FS}(a). When we turn on the Zeeman field, both Fermi surfaces are distorted and no longer possess reflection symmetry about the $k_z=0$ plane, as can be clearly seen in Fig.~\ref{FS}(b). In this perspective, the ground state is associated with nonzero total momentum along the $k_z$-axis.

Next we consider the attractive $s$-wave contact interaction between
un-like spins which, in terms of the creation and annihilation field operators
for the original spin states, is represented by 
\begin{equation}
\mathcal{H}_{{\rm int}}=U_{0}\int d{\bf r}\,\psi_{\uparrow}^{\dagger}({\bf r})\psi_{\downarrow}^{\dagger}({\bf r})\psi_{\downarrow}({\bf r})\psi_{\uparrow}({\bf r})\label{eq:swave}
\end{equation}
where $U_0$ is the bare coupling strength to be renormalized using
the $s$-wave scattering length $a_s$. In this work, we constrain our attention to the experimentally exploited broad Feshbach resonances, which is well captured by the single-channel Hamiltonian prescribed above.

\subsection{Functional Path Integral Formalism}

In this section, we briefly outline the functional path integral technique
\cite{sademelo, hu2, Stoof} and start from the partition
function $
\mathcal{Z}=\int\mathcal{D}[\psi\left(\mathbf{r},\tau\right),\bar{\psi}\left(\mathbf{r},\tau\right)]\exp\left\{ -S\left[\psi\left(\mathbf{r},\tau\right),\bar{\psi}\left(\mathbf{r},\tau\right)\right]\right\} $
where the action 
\begin{equation}
S\left[\psi,\bar{\psi}\right]=\int_{0}^{\beta}d\tau\left[\int d {\bf r} \sum_{\sigma}\bar{\psi}_{\sigma}\left(\mathbf{r},
\tau\right)\partial_{\tau}\psi_{\sigma}\left(\mathbf{r},\tau\right)+\mathcal{H}\left(\psi,\bar{\psi}\right)\right] \end{equation}
is written as an integral over imaginary time $\tau$. Here $\beta=1/(k_{B}T)$
is the inverse temperature and $\mathcal{H}\left(\psi,\bar{\psi}\right)$
is obtained by replacing  field operators $\psi^{\dagger}$ and
$\psi$ with  grassmann variables $\bar{\psi}$ and $\psi$, respectively.
We can integrate out the quartic interaction term using the Hubbard-Stratonovich
transformation \cite{Stoof},
%\bw
%\[
%e^{-U_{0}\int dxd\tau\bar{\psi}_{\uparrow}\bar{\psi}_{\downarrow}\psi_{\downarrow}\psi_{\uparrow}}=\int\mathcal{D}\left[\Delta,\bar{\Delta}\right]\exp\left\{ \int_{0}^{\beta}d\tau\int d\mathbf{r}\left[\frac{\left\vert \Delta\left(\mathbf{r},\tau\right)\right\vert ^{2}}{U_{0}}+\left(\bar{\Delta}\psi_{\downarrow}\psi_{\uparrow}\mathbf{+}\Delta\bar{\psi}_{\uparrow}\bar{\psi}_{\downarrow}\right)\right]\right\} \]
%\ew
from which the pairing field $\Delta\left(\mathbf{r},\tau\right)$
is defined. 
If we assume the mean-field order parameter to be of FF-type $\Delta=\Delta_{0}e^{i{\bf q}\cdot{\bf r}}$,
and further integrate out the fermionic fields, we arrive at an effective action as
\begin{eqnarray}
S_{\mathrm{eff}}=\int_{0}^{\beta}d\tau\int d{\bf r}(-\frac{|\Delta|^{2}}{U_{0}})-\frac{1}{2}\mathrm{Tr}\log[-\mathcal{G}_{\Delta}^{-1}]+\beta\sum_{{\bf k}}\frac{\xi_{{\bf k}+{\bf q}/2}+\xi_{-{\bf k}+{\bf q}/2}}{2}\label{eq:Seff}, \\
%\end{equation}
%\ew where \bw
%\begin{equation}
%\mathcal{G}_{\Delta}^{-1}({\bf k},i\omega_{m})=\left[\begin{array}{cc}
%i\omega_{m}-\xi_{\mathbf{k}+{\bf q}/2}-\sum_{i}\left(v_{i}(k_{i}+\frac{q_{i}}{2})+\Lambda_{i}\right)\sigma_{i} & i\Delta_{0}\hat{\sigma}_{y}\\
%-i\Delta_{0}\hat{\sigma}_{y} & i\omega_{m}+\xi_{\mathbf{k}-{\bf q}/2}+\sum_{i}\left(-v_{i}(k_{i}-\frac{q_{i}}{2})+\Lambda_{i}\right)\sigma_{i}\end{array}\right]\label{eq:Ggapinv}
\mathcal{G}_{\Delta}^{-1}({\bf k},i\omega_{m})=\left[\begin{array}{cc}
i\omega_{m}-\xi_{\mathbf{k}+{\bf q}/2}-f_+ & i\Delta_{0}\hat{\sigma}_{y}\\
-i\Delta_{0}\hat{\sigma}_{y} & i\omega_{m}+\xi_{\mathbf{k}-{\bf q}/2}-f_- \end{array}\right]\label{eq:Ggapinv}
\end{eqnarray}
where $f_\pm= \sum_{i}\left(v_{i}(k_{i} \pm \frac{q_{i}}{2}) \pm \Lambda_{i}\right)\sigma_{i}$.
In the second term of Eq.~(\ref{eq:Seff}), the trace is to be taken
over the Nambu spinor space $\Phi\left(\mathbf{r,}\tau\right)\equiv[\psi_{\uparrow},\psi_{\downarrow},\bar{\psi}_{\uparrow},\bar{\psi}_{\downarrow}]^{T}$,
the real coordinate space and imaginary time. The last term in Eq.~(\ref{eq:Seff})
comes from interchanging fermionic fields $\bar{\psi}_{\uparrow}$
and $\bar{\psi}_{\downarrow}$with $\psi_{\uparrow}$ and $\psi_{\downarrow}$
and the corresponding equal-time limiting procedure \cite{Stoof}.
From Eq.~(\ref{eq:Seff}), we can further sum over Matsubara frequencies
to arrive at the grand thermodynamic potential: 
\begin{eqnarray}
\frac{\Omega}{V} & = & -\frac{1}{\beta}\ln\mathcal{Z}
  =  -\frac{|\Delta|^{2}m}{4\pi\hbar^{2}a_{s}}+
 \frac{1}{V}\sum_{\bf k}\left[\frac{\xi_{{\bf k}+{\bf q}/2}+\xi_{-{\bf k}+{\bf q}/2}}{2}-\frac{1}{4}\sum_{\alpha=1}^{4}|E_{{\bf k}}^{\alpha}| \right. \nonumber \\ && \left.+\frac{|\Delta|^{2}}{2\epsilon_{{\bf k}}}-\frac{1}{2\beta}\sum_{\alpha=1}^{4}\ln \Big(1+\exp(-\beta|E_{{\bf k}}^{\alpha}|)\Big)\right] \label{omega}
\end{eqnarray}
where we have regularized the bare interaction strength $U_{0}$ in terms
of the $s$-wave scattering length $a_{s}$ by 
$\frac{1}{U_{0}}=\frac{m}{4\pi\hbar^{2}a_{s}}-\frac{1}{V}\sum_{{\bf k}}\frac{1}{2\epsilon_{{\bf k}}}$. $E_{{\bf k}}^{\alpha}$ ($\alpha=1$, 2, 3, 4) are the quasi-particle energy dispersion, which are just the four eigenvalues obtained by solving
$\det[\mathcal{G}_{\Delta}^{-1}({\bf k},E_{{\bf k}}^{\alpha})]=0$.
In our case, $E_{{\bf k}}^{\alpha}$ are too complicated to be
presented here.%, and we put further derivations in Appendix A. 

In the following sections, we restrain our attention to the isotropic 3DSOC with Zeeman field along the $z$-axis,
i.e., ${\bf \Lambda}=(0,\,0,\, h)$, and use the ansatz for FF
order parameter $\Delta({\bf r})=\Delta e^{iqz}$. 
%We shall justify the validity of this ansatz later. By setting $\hbar=2m=k_B=1$, we use $E_F$, $1/k_F$ and $T_F$ as energy, length, and temperature unit. 
%In the
%following sections, we discuss about this particular type of SO coupling
%scheme and its corresponding big advantage over others \cite{Chuanweiab,yiwei,
%HuiNist}.

\begin{figure}
\begin{center}
\includegraphics[width=.68\textwidth]{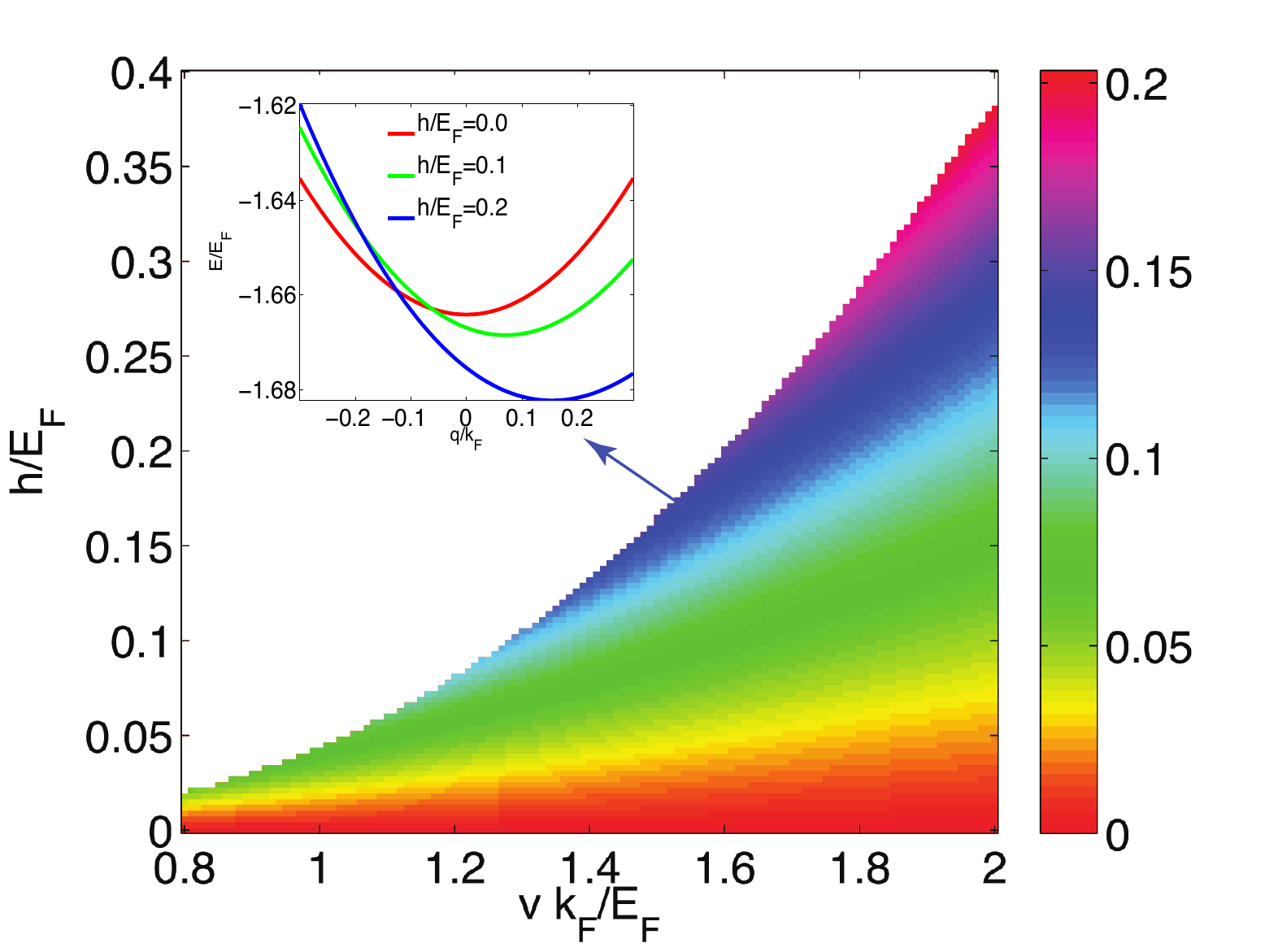}
\end{center}
\caption{Finite momentum dimer bound state solution for 3DSOC. The coloring shows the magnitude of $q_{\rm 2b}$, varying with the SO coupling strength $v$ and the Zeeman field strength $h$; the inset shows bound state energy $E_{{\bf q}}$ as a function
of $q/k_F$ (along the $z$-axis) for different $h$, from top
to bottom $h/E_{F}=0, 0.1, 0.2$. We fix scattering length as $1/(k_{F}a_{s})=-1$} \label{fig2}
\end{figure}

\section{Results on two-body problem }\label{resultsa}

Following the path integral approach, we can characterize two-body properties at low-energy sector by inverse vertex function, which we refer to \cite{leipralong} for more details. %list in Appendix B for more details. 
We found consistent results reported in our previous
paper \cite{2body} which are obtained by solving the two-body Schr{\"{o}}dinger
equation. For a bound-state with total momentum ${\bf q}$, the corresponding energy $E_{\bf q}$ is obtained by solving the following equation:
\begin{equation}
\frac{m}{4\pi\hbar^{2}a_{s}}  = \frac{1}{V}\sum_{{\bf k}}\left\{ \frac{1}{{\cal E}_{{\bf k,q}}-\frac{4v^{2}}{{\cal E}_{{\bf k,q}}}\left(\frac{{\cal E}^2_{\bf k,q}}{{\cal E}_{{\bf k,q}}^{2}-(2h+vq)^{2}}k_{\perp}^{2}+k_{z}^{2}\right)}+\frac{1}{2\epsilon_{{\bf k}}}\right\} \label{2b}
\end{equation}
where ${\cal E}_{{\bf k,q}}=E_{{\bf q}}-\epsilon_{\frac{{\bf q}}{2}+{\bf k}}-\epsilon_{\frac{{\bf q}}{2}-{\bf k}}$ and $\epsilon_{\bf k} = \hbar^2 k^2/(2m)$.
For a given set of parameters $h$, $v$ and $a_{s}$, we can numerically obtain the eigenenergy of the dimer bound state $E_{{\bf q}}$ as a function of ${\bf q}$. The momentum ${\bf q}_{0}$ at which $E_{{\bf q}}$ reaches the minimum labels the dimer state with lowest energy. The binding energy is defined as $\epsilon_{b}=2E_{\rm {min}}-E_{{\bf q}_{0}}$, where $E_{\rm min}$ is the ground state energy of single-particle spectrum $E^-({\bf k})$. Only when $\epsilon_{b}>0$ can we consider the dimer as a true two-body bound state. Otherwise, its energy lies in the single particle continuum. For the convenience of further comparison with many-body state, here we take the laser recoil momentum $k_r$, which determines the SO coupling strength, to be equal to Fermi momentum $k_F$, which is determined by typical atomic density in experiments \cite{lin, Zhang, mit}.

For a Zeeman field along the $z$-axis, we have ${\bf q}_0=q_{\rm 2b} \hat{z}$. Following the above-mentioned protocol, we plot $q_{\rm 2b}$ as a function of the SO coupling strength and the Zeeman field strength $h$ in Fig.~\ref{fig2}. As one would expect, Zeeman field tends to destroy two-body bound state; whereas SO coupling enhances its formation. The competition between these two outlines the critical boundary value, beyond which $\epsilon_b$ becomes negative and no stable bound state can be found. With increasing $h$, the minimum of $E_{{\bf q}}$
deviates further away from zero momentum to some finite value. As long as Zeeman field is \emph{non-zero}, the lowest-energy bound state would occur at \emph{finite} center-of-mass momentum $q_{\rm 2b}$. Our calculation shows that the magnitude of $q_{\rm 2b}$
can be as high as $0.2k_{F}$. 
%At first sight, this result is rather
%surprising since the $s$-wave interaction Eq.~(\ref{eq:swave}) conserves momentum during two particle scattering event. However, because of the prominent feature of SO coupling, atom's orbital motion is coupled to its (pseudo-)spin, and the $s$-wave interaction induces redistribution of spin-up and spin-down population and hence shifts the total momentum of the system. Furthermore, the existence of finite momentum dimer bound state may be regarded as a consequence of broken Galilean invariance, which is responsible for a variety of peculiar phenomena in quantum gases with SO coupling \cite{deviatedipole, ram, wubiao}. 

\section{Results on many-body problem}\label{resultsb}
Motivated by the two-body results, one naturally attempts to explore the direct analog for the many-body system, which we study in this section.

We take a canonical ensemble approach by considering a homogeneous system with fixed particle number $N$ and volume $V$, and hence the density $n=N/V=k_F^3/(3\pi^2)$. The important quantity that determines the mean field phase diagram
shall be the free energy, also known
as the Landau potential, defined as $F=\Omega+\mu N$. At zero temperature, it coincides with the ground state energy.  For a given set
of parameters (including SO coupling strength $v$,
Zeeman field strength $h$, interaction parameter $1/(k_{F}a_{s})$,
and temperature $T$), order parameter $\Delta$, chemical
potential $\mu$, and the FF momentum ${\bf q}=q_{\rm FF}\,\hat{z}$ should be determined self-consistently by stationary conditions
\begin{equation}
\frac{\partial F}{\partial\Delta}=0,\;\;\frac{\partial F}{\partial\mu}=0,\;\;\frac{\partial F}{\partial{ q}}=0.\label{eq:stationarycondition}
\end{equation}
We shall
explicitly consider three types of phase: normal gas ($\Delta=0$,
${ q}=0$), BCS state ($\Delta\neq0$, ${ q}=0$), and FF state
($\Delta \neq 0$, ${ q}_{\rm FF} \neq 0$).  

\begin{figure}
\begin{center}
\includegraphics[width=.68\textwidth]{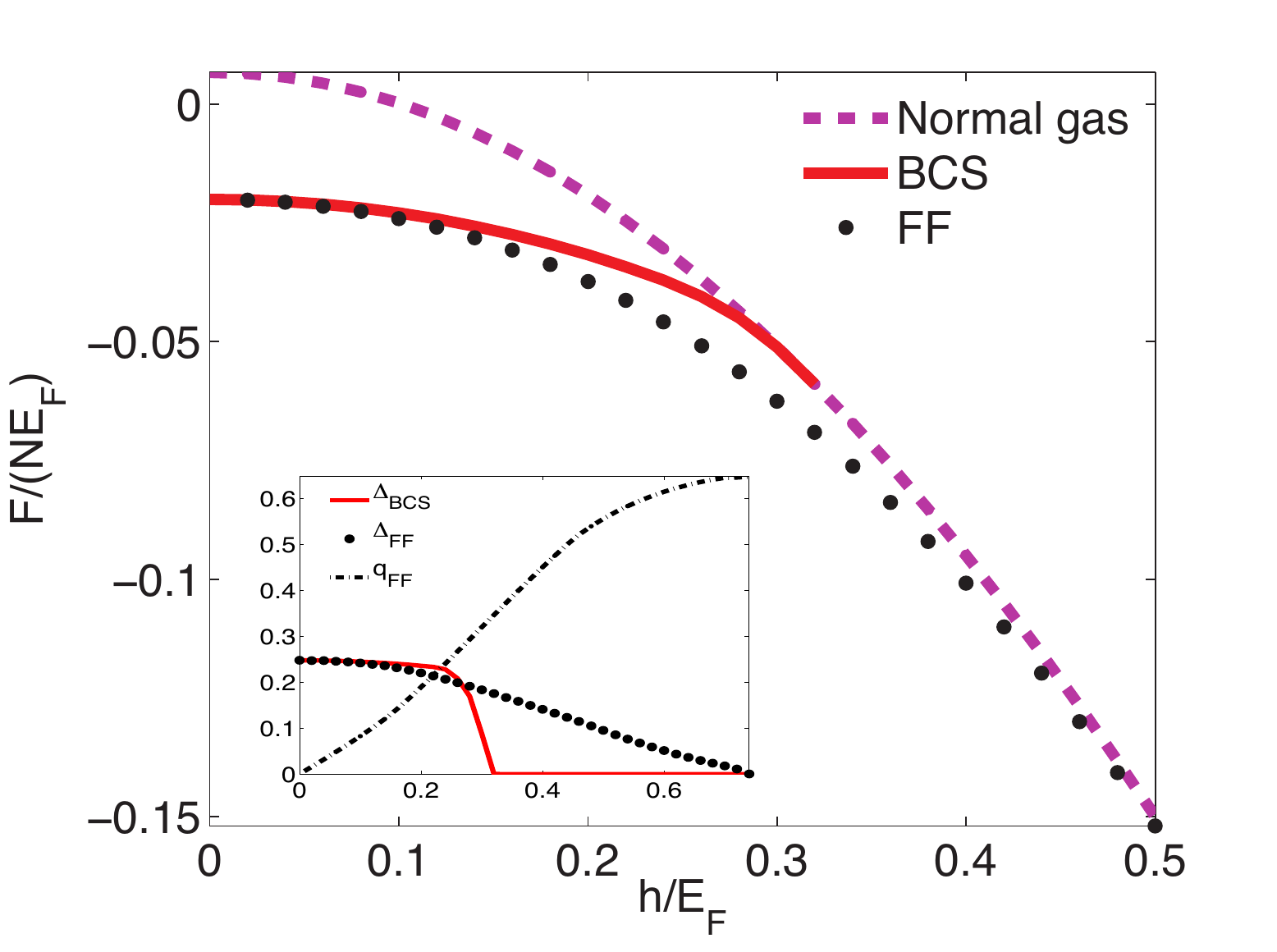}
\end{center}
\caption{Zero temperature free energy as a function of $h$ for fixed
SO coupling strength $v=E_F/k_F$ and interaction strength $1/k_Fa_s=-1$. FF superfluid phase is favored throughout the plotted parameter space. %At large Zeeman field limit, it connects to normal %phase smoothly. 
In the inset, we plot the BCS order parameter $\Delta_{\rm BCS}$, the FF order parameter $\Delta_{\rm FF}$ (both in units of $E_F$), and the FF momentum $q_{\rm FF}$ (in units of $k_F$) as functions of $h$. } \label{fig3} 
\end{figure}

\subsection{Zero-temperature phase diagram on the BCS side}
We shall first focus on a relatively weak-interacting 
regime on the BCS side of crossover and take $1/(k_{F}a_{s})=-1$. In this regime, we can easily justify the mean-field treatment at both zero and finite temperature, and furthermore the SO coupling
effect would be more pronounced \cite{YuZhai,
Iskin, GongZhang, HuiPu}. 

To get some insights first, in Fig.~\ref{fig3},
we plot free energy as a function of $h$ for a given SO coupling strength $v=E_F/k_F$. (We choose this relatively large SO strength, to avoid possible complications, e.g. Sarma phase \cite{hui06rapid}, phase separation \cite{leocomment} etc.)
It is very remarkable to notice that FF state is energetically favored
for arbitrarily small $h$. For instance, at $h=0.02E_F$,  the gain of energy over the BCS pairing phase is $\Delta F=F_{\rm {BCS}}-F_{\rm {FF}}\approx4.54588\times10^{-5}NE_F$. However, this energy gain quickly increases as $h$ is increased. For example, at $h=0.28E_F$, we have $\Delta F=1.13728\times 10^{-2}NE_F$ which is more than two orders of magnitude larger and represents a very large energy value on the BCS side of Feshbach resonance. Once again, the idea of favoring FF phase is backed by the picture of Fermi surface deformation (cf. Fig.~\ref{FS}) and two-body bound state solutions [Eq.~(\ref{2b}) and Fig.~\ref{fig2}]. When we further increase $h$, BCS superfluid is taken over by normal phase as the BCS order parameter drops to zero rather sharply (see the inset of Fig.~\ref{fig3}); on the other hand, FF state connects to normal phase very smoothly
at a much larger value of $h$. 
%For instance, at the data point $h=0.5E_F$, $F_{\text{Normal}}-F_{\text{FF}}\approx1.979930\times10^{-3}NE_F$. 

\begin{figure}
\begin{center}
\includegraphics[width=.88\textwidth]{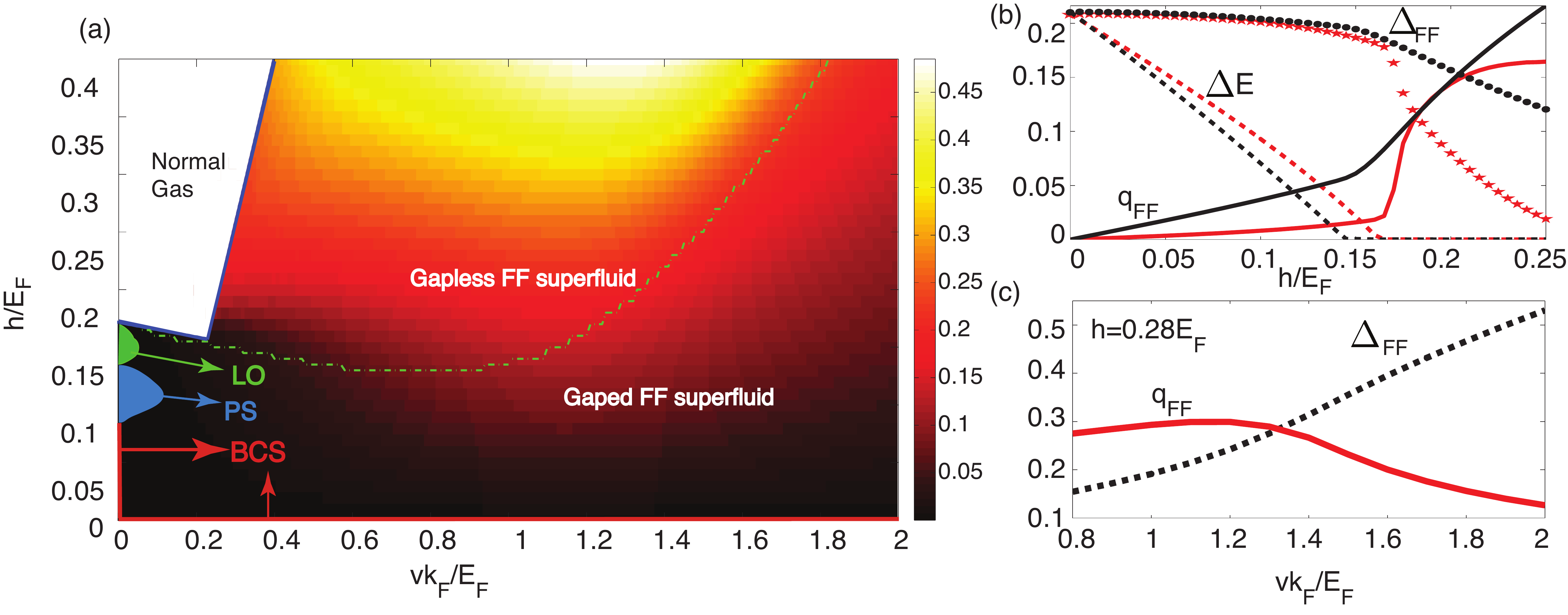}%{zeroTphasediagram2psai}
\end{center}
\caption{(a) Zero-temperature phase diagram at $1/k_Fa_s=-1$ in the parameter space spanned by $h$ and $v$. %BCS phase is not stable against FF pairing and it only exists within the axis. In the absence of SO coupling, with increasing %Zeeman field, LO phase emerges within a narrow window; however, in the presence of SO coupling,  FF phase is more favorable, %and excitation energy under goes gapped to gapless transition. 
The FF phase is divided into gapped and gapless region by the green dashed line. 
%The critical value of $h$ is not very sensitive to coupling strength.
%is the continuation of BCS (strictly on axis only) and it is distinctively different from
%gapless LO phase. 
%Note that the mean field result is consistent with two body calculation that as long as Zeeman field is \emph{non-zero}, in %principle, the bound state would favor lower energy at \emph{finite} center-of-mass momentum. 
BCS state only exists strictly on the axis marked by two red straight lines. The two blue lines indicate the smooth boundary between FF state and normal phase. Within the FF phase, the color scale indicates the momentum $q_{\rm FF}$. LO and phase separation (PS) regions are added schematically for illustration purpose.
(b) Single-particle excitation gap $\Delta E$, FF order parameter $\Delta_{\rm FF}$ and momentum $q_{\rm FF}$ as functions of $h$. The SO coupling strengths are $v=0.2E_F/k_F$ (red curves) and $0.5 E_F/k_F$ (black curves). 
(c) $\Delta_{\rm FF}$ and $q_{\rm FF}$ as functions of $v$ for $h=0.28E_F$. In all plots, the energy is in units of $E_F$, and momentum in units of $k_F$.}
\label{fig4}
\end{figure}

The FF state here has different origin with the conventional FFLO
states in the absence of the SO coupling \cite{Orso, Hui2007, Ueda08}, in which case, for a
given interaction strength and with increasing population imbalance (i.e., Zeeman field),
one would expect that competitions among various quantum phases (BCS,
Sarma, FFLO, and normal phases) could lead to both first- and second-order phase transitions. By contrast in the presence of
SO coupling, especially 3DSOC, FF state dominates almost the entire phase diagram, as we map out in the $v$-$h$ plane Fig.~\ref{fig4}(a). The BCS phase only exists on the axis (i.e., in the absence of either the Zeeman field or the SO coupling). Normal phase and FF phase are connected by a smooth boundary, which we identify by setting a threshold value of energy difference $|F_{\rm FF} - F_{\rm normal}| \approx 10^{-5}NE_F$. Note that close to the boundary, $\Delta_{{\rm FF}}$ also becomes exceedingly small. For illustration purposes, we schematically added two small regions near $v=0$, the LO (green) and the phase separation (blue) regions, in the phase diagram. The boundaries of these two phases in the absence of the SO coupling (i.e., at $v=0$), which are well studied, are obtained from previous results  \cite{hui06rapid,leocomment, liuwm}. It has been shown that, with increasing SO coupling strength, both these phases are suppressed rather rapidly \cite{liuwm}.  
%The small occupation of LO state can be understood from time-reversal symmetry breaking caused by Zeeman field. Out of LO and FF pairing possibilities, albeit being both finite-momentum pairs, $+{\bf q}$ and $-{\bf q}$ are no longer symmetric, thus the odds of LO channel is suppressed.

Furthermore, in Fig.~\ref{fig4}(a), FF phase is divided into the gapped and the gapless regions by examining the single-particle excitation gap $\Delta E= \min\{|E_{{\bf k}}^{\alpha}|\}$, where $E_{{\bf k}}^{\alpha}$ are quasi-particle dispersions introduced in Eq.~(\ref{omega}). As shown in Fig.~\ref{fig4}(b), $\Delta E$ decreases monotonically as a function of $h$ and drops to zero at some critical value of $h_c$ which depends on the SO coupling strength $v$. The critical value $h_c$ is represented by the green dashed line in Fig.~\ref{fig4}(a). At $h_c$, both $q_{\rm {FF}}$ and $\Delta_{\rm {FF}}$ exhibit kinks for relatively small SO coupling strength. These kinks get washed out quickly with increasing $v$ (see, for instance, the inset of Fig.~\ref{fig3}). On the other hand, in the limit of $v=0$, these kinks become true jumps signaling the first-order phase transition between the BCS phase and the FF phase region. In Fig.~\ref{fig4}(c), we plot $q_{\rm FF}$ and $\Delta_{\rm FF}$ as functions of $v$ for a fixed $h$. We note that even though $\Delta_{\rm FF}$ increases monotonically as $v$, the FF momentum $q_{\rm FF}$ shows non-monotonic behavior: it first increases and then decreases as $v$ is increased from zero. 

\begin{figure}
\begin{center}
\includegraphics[width=.58\textwidth]{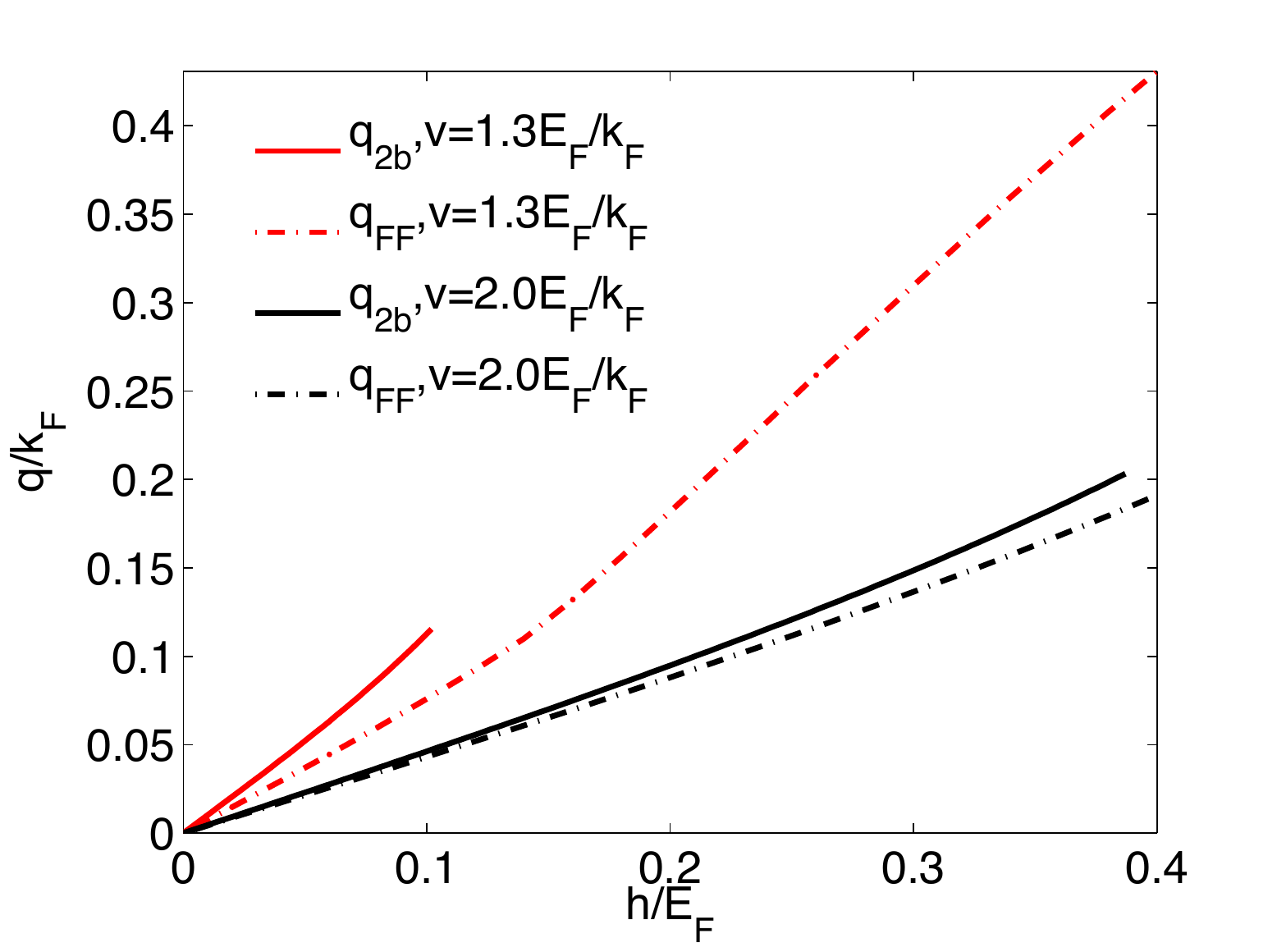}
\end{center}
\caption{Momentum comparison of the FF state and two-body bound state at $1/k_Fa_s=-1$.  }
\label{fig7}
\end{figure}

It is instructive to make comparisons between the two-body results and the many-body results. To this end, we consider a cloud of degenerate Fermi gas
typically realized in experiment, with density $n=10^{12} {\rm cm}^{-3}$
which defines $k_F$ and $E_{F}$. We compare 
the two-body dimer momentum $q_{\rm 2b}$ with the many-body FF
pairing momentum $q_{\rm FF}$ in Fig.~\ref{fig7} at two different values of SO coupling strengths. Note that the range of $h$ values for which the two-body bound state exists is much smaller than that for the existence of the FF state. For example, at $v=1.3 E_F/k_F$, two-body bound states only exist for $h<0.1E_F$; while the FF state extends all the way up to about $h \approx E_F$. As such, the largest $q_{\rm FF}$ that can be achieved is much larger than the largest $q_{\rm 2b}$. In the region where both two-body bound state and the FF state exist, $q_{\rm FF}$ and $q_{\rm 2b}$ are comparable, with the latter somewhat larger. The difference between them, however, becomes smaller as the SO coupling strength increases, indicating that at large SO coupling strength, the many-body properties of the system are also dominated by the two-body physics. 
%the exist condition of  two body bound state ($\epsilon_b>0$) does not necessarily coincide with the exist condition of superfluid state ($|\Delta_{\text{FF}}|>0$). In general, many bound superfluid state is more robust against Zeeman field, as seen in Fig.~\ref{fig7}. For coupling strength $v=1.3E_F/k_F$, with increasing Zeeman field strength, two-body bound state ceases to exist around $h=0.1E_F$, whereas FF superfluid state continues to survive. For the overlap part, difference between dimer state momentum $q_{\text{2b}}$ and FF momentum $q_{\text{FF}}$ becomes smaller when coupling strength becomes larger, say $v=2E_F/k_F$. This result comes as expected because in the limit of large coupling strength, two body effect dominates in the many body system, and  $q_{\text{2b}}\approx q_{\text{FF}}$ is just one of the manifestation.  

\begin{figure}
\begin{center}
\includegraphics[width=.78\textwidth]{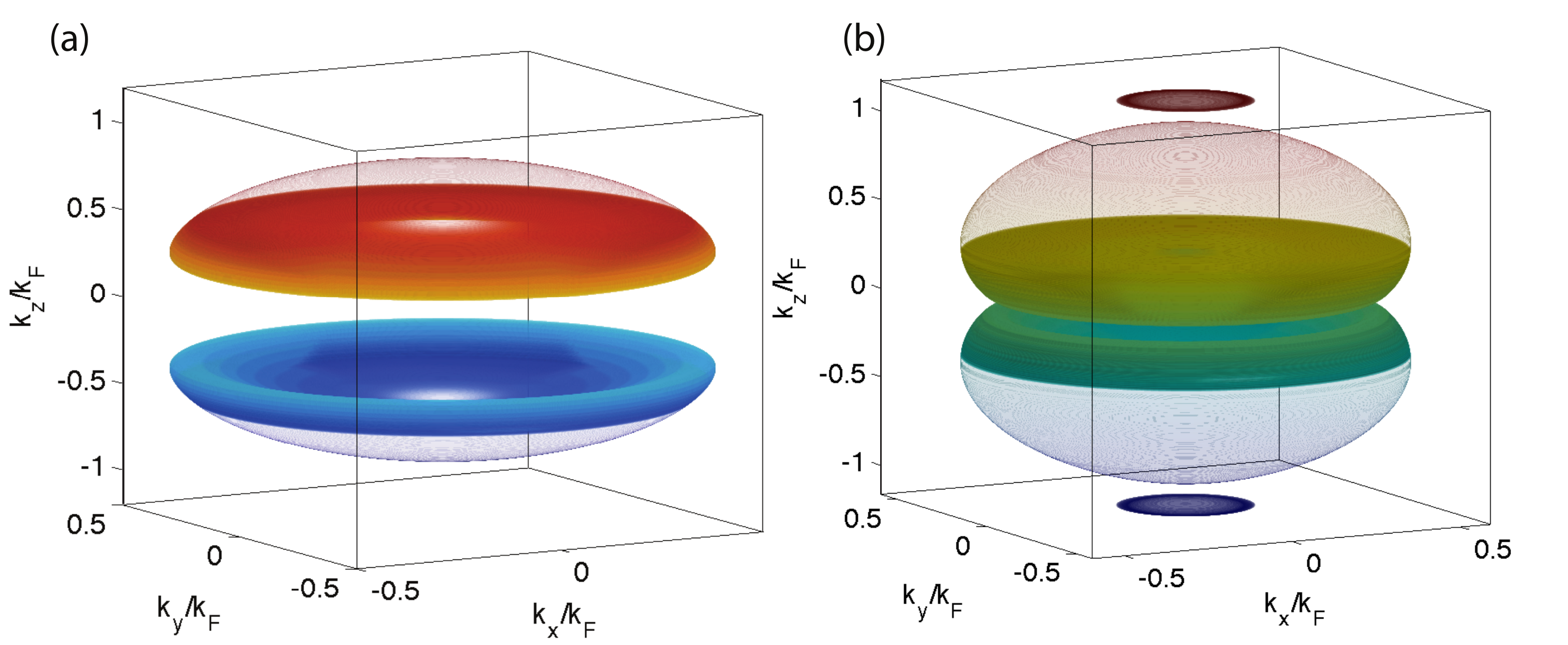}
\end{center}
\caption{Nodal Fermi surface plots in momentum space. The closed surfaces are formed by momentum values at which the excitation gap $\Delta E$ vanishes. Both figures have the same interaction parameter $1/k_Fa_s=-1$ as in phase diagram Fig.~\ref{fig3} and coupling strength $v=0.5E_F/k_F$, while $h=0.2E_F$ for (a) and $h=0.35 E_F$ for (b). }\label{NFS}
\end{figure}

Before ending this subsection, we want to remark on the gapless FF state. For Zeeman field strength above the critical value $h_c$, one or more quasi-particle energy $E_{\bf k}^\alpha$ will vanish at certain values of momentum ${\bf k}$. Such momenta form closed surfaces (nodal Fermi surface) in momentum space with cylindrical symmetry around the $k_z$-axis and reflection symmetry about the $k_z=0$ plane. Hence such nodal Fermi surfaces always appear in pairs and may be measured using the technique of momentum-resolved radio-frequency spectroscopy. Two examples are illustrated in Fig.~\ref{NFS}.

%The other issue one might worry is the phase separation regime, a
%situation where homogeneous condition is no longer satisfied. To address
%this concern, we adopt grand canonical approach by fixing chemical
%potential and plot out $\Omega$ in phase space. At least we can say
%for sure that in the large limit of coupling strength, the single
%minimum of FF state dominates significantly. 
%\begin{figure}
%\includegraphics[width=.48\textwidth]{grand_omega_large}\caption{(Color Online) to be modified or deleted.}
%\end{figure}

\subsection{Effects of interaction}\label{interaction}

\begin{figure}
\begin{center}
\includegraphics[width=.68\textwidth]{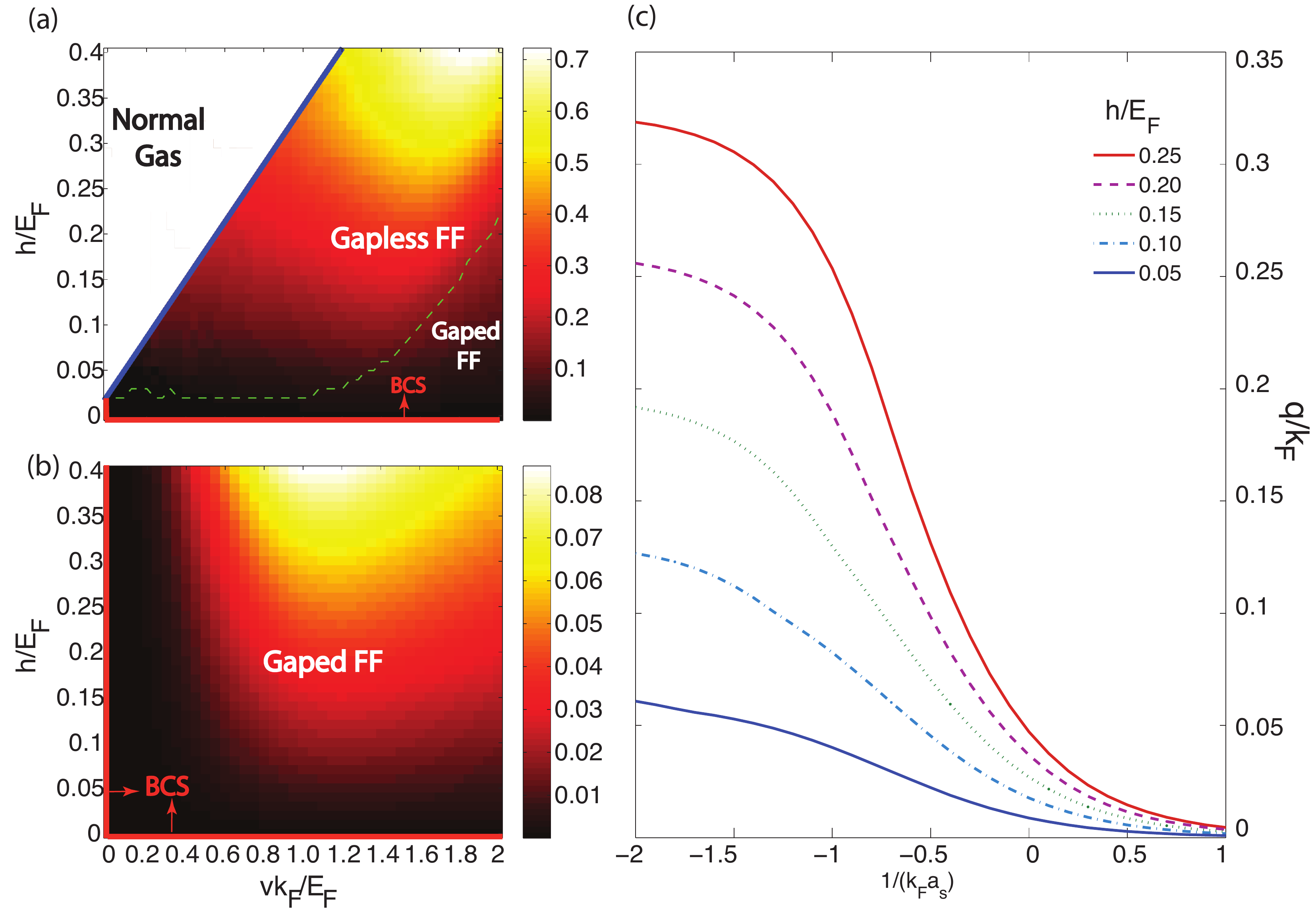}
\end{center}
\caption{Zero temperature phase diagram with interaction strength $1/k_F a_s=-2$ and $1/k_F a_s=0$ for (a) and (b), respectively. The color scale represents $q_{\rm FF}/k_F$. (c) FF superfluid momentum $q_{\rm FF}$ as a function of the interaction strength. For curves from top to bottom, $h/E_F=0.25,~0.2,~0.15,~0.1,~0.05$, and the SO coupling strength is fixed to be $v=E_F/k_F$.}
\label{fig8}
\end{figure}

So far we have focused on the zero-temperature phase diagram of a weakly-interacting system. Now we briefly discuss the effects of interaction in this subsection and those of finite temperature in the next.
In Fig.~\ref{fig8}(a) and (b), we present two zero-temperature phase diagrams in the $h$-$v$ plane for $1/k_F a_s= -2$ and $0$, respectively. They are qualitatively similar to the one presented in Fig.~\ref{fig4}(a) for $1/k_F a_s=-1$.  As we move from the BCS limit towards unitarity, the region of normal phase shrinks and the FF superfluid remains dominant. Furthermore, the region for gapped FF phase increases quickly. At unitarity, the whole parameter space presented in Fig.~\ref{fig8}(b) are occupied by the gapped FF phase. On the other hand, for fixed $h$ and $v$, the FF momentum $q_{\rm FF}$ quickly decreases as we go from the BCS side to the BEC side, as shown in Fig.~\ref{fig8}(c). This result is consistent with the one obtained from the two-body study \cite{2body,2body1}. 

%\begin{figure}
%\begin{center}
%\includegraphics[width=.88\textwidth]{BCS_unitary}
%\end{center}
%\caption{(Color Online) effect of interaction.}
%\label{fig9}
%\end{figure}

\subsection{Effects of temperature}\label{temperature}

\begin{figure}
\begin{center}
\includegraphics[width=.88\textwidth]{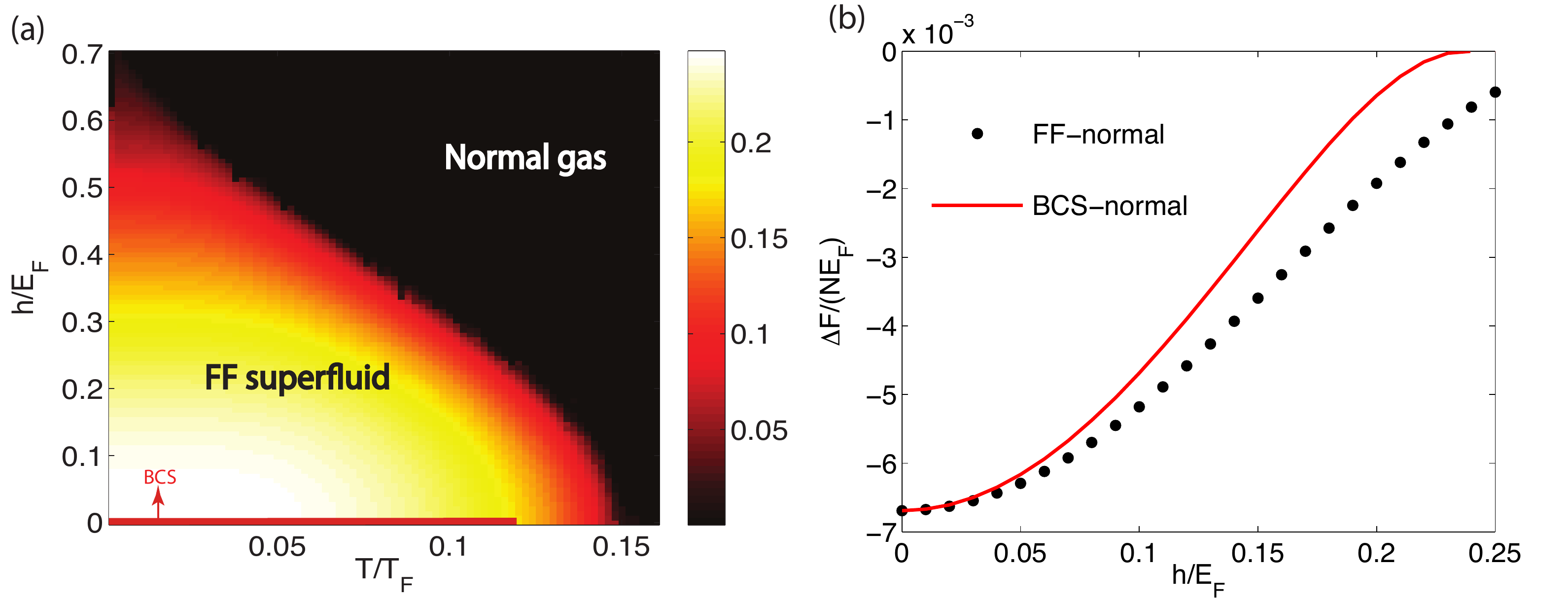}
\end{center}
\caption{(a) Finite temperature phase diagram at $1/k_F a_s=-1$ and $v=E_F/k_F$. The color scale indicates FF order parameter $\Delta_{{\rm FF}}$. (b) The free energy difference between the two superfluid phases (BCS and FF) and the normal phase at temperature $T=0.1T_F$. FF phase always has lowest free energy. }\label{finiteT}
\end{figure}

Finally, we consider the effects of finite temperature. In Fig.~\ref{finiteT}(a), we plot the phase diagram in the parameter space spanned by $h$ and $T$ by taking $1/(k_Fa_s)=-1$ and $v=E_F/k_F$. The FF superfluid phase dominates at small $h$ and low $T$. There is a second order transition towards normal phase as $h$ and/or $T$ increases. The BCS phase again only lives on the $h=0$ axis. In Fig.~\ref{finiteT}(b), we compare the free energies for all three phases at $T=0.1T_F$ and clearly show that the FF phase possesses the lowest free energy at any finite values of $h$ as long as $h$ is below a threshold at which the system turns normal.

%
%\begin{figure}
%\begin{tabular}{|c|c|c|}
%\hline 
%\includegraphics[scale=0.2]{T_h_0\lyxdot 1_v_0\lyxdot 3_d_v2} & \includegraphics[scale=0.2]{T_h_0\lyxdot 1_v_0\lyxdot 6_d_v2} & \includegraphics[scale=0.2]{T_h_0\lyxdot 18_v_0\lyxdot 6_d_v2}\tabularnewline
%\hline
%\end{tabular}
%
%\caption{(Color online) Finite temperature FF state. From left to right, $h/E_{F}$
%is $0.1,0.18,0.1$ and $\lambda k_{F}/E_{F}$ is $0.3,0.6,0.6$ }
% Now for the sake of completeness, we consider effect of temperature. We first start from 
%\begin{figure}[htp]
%\includegraphics[width=.48\textwidth]{}\caption{(Color Online) }
%\end{figure}

%
%\end{figure}
\section{Conclusion}\label{Conclusion}
In summary, we have studied spin-orbit coupled Fermi gas subjected to an effective Zeeman field. Based on the picture of Fermi surface deformation, the two-body calculations, and the mean-field many-body results, we conclude that the BCS state with zero-momentum Cooper pairs is not stable against Fulde-Ferrell superfluid pairing at any finite Zeeman field strength. The FF phase is robust against interaction and finite temperature, and the corresponding center-of-mass momentum of the Cooper pair can be comparable to the Fermi momentum. The finite-momentum dimer state in the two-body situation and the FF state in the many-body setting both originate from the asymmetric momentum distribution as a consequence of the interplay between spin-orbit coupling and the Zeeman field. This asymmetry also determines the direction of the momentum for the dimer state or the Cooper pairs. In this sense, the FF state we discussed in the present work is not exactly the same as the FF or FFLO state in the context of a spin-imbalanced Fermi gas without spin-orbit coupling. In the latter case, the direction of the momentum of the Cooper pairs is determined through the mechanism of spontaneous symmetry breaking. For a similar reason, we only considered FF state in our work, not the FFLO state. The FFLO state requires the Cooper pairs to possess two momenta with equal magnitude but opposite directions. However, in our case, the asymmetric momentum distribution uniquely picks one particular momentum rather than an opposite pair. 

\ack 
HP is supported by the NSF, the Welch Foundation (Grant No. C-1669), and the DARPA OLE program. We would like to thank Hui Hu and Xia-ji Liu for stimulating discussions, Xiangfa Zhou and Wei Yi for helpful comments, and B. Ramachandhran for the help on improving the quality of some figures. We thank Wei Yi for sending us their preprint \cite{yi} in which they investigated the same system from a grand canonical ensemble approach. The essential results of our works agree with each other.


\begin{thebibliography}{99}
\bibitem{BCS} J. Bardeeen, L.N. Cooper, and J. R. Schrieffer, Phys. Rev. {\bf 106}, 162 (1957) and Phys. Rev. {\bf 108}, 1175 (1957) .

\bibitem{sed05} Sedrakian, A., Mur-Petit, J., Polls, A. and Muther, H. , Phys. Rev. A {\bf 72}, 013613 (2005).

\bibitem{FF} P. Fulde, and R. A. Ferrell, Phys. Rev. {\bf 135}, A550 (1964).

\bibitem{LO} A. I. Larkin and Y. N. Ovchinnikov, Sov. Phys. JETP {\bf 20}, 762 (1965).

\bibitem{FFLOrmpQCD} R. Casalbuoni and G. Nardulli, Rev. Mod. Phys. {\bf 76}, 263320 (2004).  

\bibitem{kun}K. Yang, Phys. Rev. B {\bf 63}, 140511(R) (2001).
\bibitem{feiguin}A. E. Feiguin, and F. Heidrich-Meisner, Phys. Rev. B {\bf 76}, 220508(R) (2007).
\bibitem{1d}A. L\"{u}scher, R. M. Noack, and A. M. L\"{a}uchli, Phys. Rev. A {\bf 78}, 013637 (2008).
\bibitem{erich}M. Casula, D. M. Ceperley, E. J. Mueller, Phys. Rev. A {\bf 78}, 033607 (2008).

\bibitem{Orso} G. Orso, Phys. Rev. Lett. {\bf 98} 070402 (2007).

\bibitem{Hui2007} H. Hu, X.-J. Liu, P. D. Drummond, Phys. Rev. Lett. {\bf 98} 070403 (2007).

\bibitem{Ueda08} M. Tezuka, M. Ueda, Phys. Rev. Lett. {\bf 100} 010403 (2008) .

\bibitem{honglu} H. Lu, L. O. Baksmaty, C. J. Bolech, and H. Pu, Phys. Rev. Lett. {\bf 108}, 225302 (2012).

\bibitem{exp}C. J. Bolech, F. Heidrich-Meisner, S. Langer, I. P. McCulloch, G. Orso, and M. Rigol, Phys. Rev. Lett. {\bf 109}, 110602 (2012).
\bibitem{exp1}J. Kajala, F. Massel, and P. T\"{o}rm\"{a}, Phys. Rev. A {\bf 84}, 041601(R) (2011).

\bibitem{guan}X.-W. Guan, M. T. Batchelor, and C. Lee, arXiv:1301.6446 (2013); A. Feiguin, F. Heidrich-Meisner, G. Orso, W. Zwerger, Lect. Notes. Phys. {\bf 836}, 503 (2012).

\bibitem{1DRICE} Y.-A. Liao, A. S. C. Rittner, T. Paprotta, W. Li, G. B. Partridge, R. G. Hulet, S. K. Baur, and E. J. Mueller, Nature {\bf 467}, 567 (2010).

\bibitem{3DMIT} M. W. Zwierlein, A. Schirotzek, C. H. Schunck, and W. Ketterle, Science {\bf 311}, 492 (2006).

\bibitem{3DRICE} G. B. Partridge, W. Li, R. I. Kamar, Y.-A. Liao, and R. G. Hulet, Science {\bf 311}, 503 (2006).

\bibitem{parish07} M. M. Parish, F. M. Marchetti, A. Lamacraft, and B. D. Simons, Nature Physics {\bf 3}, 124 (2007).

\bibitem{note} However, the parameter space for FFLO ground state may be enlarged in the presence of optical lattices. See, for example, Y. L. Loh, and N. Trivedi, Phys. Rev. Lett. {\bf 104}, 165302 (2010); D. H. Kim, and P. T\"{o}rm\"{a}, Phys. Rev. B {\bf 85}, 180508(R) (2012).

\bibitem{lin} Y.-J. Lin {\em et al.}, Phys. Rev. Lett. {\bf 102}, 130401 (2009); Y.-J. Lin, R. L. Compton, K. Jimenez-Garcia, J. V. Porto and I. B. Spielman, Nature, {\bf 426}, 628 (2009); Y.-J. Lin, K. Jimenez-Garcia, and I. B. Spielman, Nature (London) {\bf 471}, 83 (2011).  Y.-J. Lin, R. L. Compton, K. Jimenez-Garcia, W. D. Phillips, J. V. Porto and I. B. Spielman, Nature Physics, {\bf 7}, 531 (2011).

\bibitem{Zhang} P. Wang, Z.-Q. Yu, Z. Fu, J. Miao, L. Huang, S. Chai, H. Zhai, and J. Zhang, Phys. Rev. Lett. {\bf 109}, 095301 (2012).

\bibitem{mit} L. W. Cheuk, A. T. Sommer, Z. Hadzibabic, T. Yefsah, W. S. Bakr, and M. W. Zwierlein, Phys. Rev. Lett. {\bf 109}, 095302 (2012).

\bibitem{2body} L. Dong, L. Jiang, H. Hu, and H. Pu, Phys. Rev. A {\bf 87}, 043616 (2013).

\bibitem{2body1}V. B. Shenoy, arXiv:1211.1831 (2012).

\bibitem{Chuanweiab} Z. Zheng, M. Gong, X. Zou, C. Zhang, and G.-C. Guo, Phys. Rev. A {\bf 87}, 031602(R) (2013), and arXiv: 1212.6826 (2012).

\bibitem{yiwei} F. Wu, G.-C. Guo, W. Zhang, and W. Yi, Phys. Rev. Lett. {\bf 110}, 110401 (2013).

\bibitem{HuiNist}  X.-J. Liu, and H. Hu, arXiv:1302.0553 (2013); H. Hu, and X.-J. Liu, arXiv:1304.0387 (2013).

\bibitem{3DSOC} B. M. Anderson, G. Juzeliunas, V. M. Galitski, and I. B. Spielman, Phys. Rev. Lett. {\bf 108}, 235301 (2012).

\bibitem{zhou} Y. Li, X. Zhou, and C. Wu, Phys. Rev. B {\bf 85}, 125122 (2012).

\bibitem{shenoyFS} J. P. Vyasanakere, S. Zhang, and V. B. Shenoy, Phys. Rev. B {\bf 84}, 014512 (2011).

%\bibitem{3Dtop} M. Z. Hasan and C. L. Kane, Rev. Mod. Phys. {\bf 82} 3045, (2010).
%
%\bibitem{weyla} X. Wan, Ari M. Turner, Ashvin Vishwanath, and Sergey Y. Savrasov, Phys. Rev. B {\bf 83}, 205101 (2011).
%
%\bibitem{weylb}  D. A. Pesin and L. Balents, Nature Phys. {\bf 6}, 376 (2010).

\bibitem{Jacob2007} A. Jacob, P. Ohberg, G. Juzeliunas, L. Santos, Appl. Phys. B. {\bf 89}, 439, (2007).

\bibitem{RMP} J. Dalibard, F. Gerbier, G. Juzeliunas, and P. Ohberg, Rev. Mod. Phys. {\bf 83}, 1523 (2011).

\bibitem{sademelo} C. A. R. Sá de Melo, M. Randeria, and J. R. Engelbrecht, Phys. Rev. Lett. {\bf 71}, 3202 (1993); M. Randeria, in \emph{Bose-Einstein Condensation}, edited by A. Griffin, D. W. Snoke, and S. Stringari, (Cambridge University Press, Cambridge, England, 1995), p. 355-392.

\bibitem{hu2} H. Hu, X.-J. Liu, and P. Drummond, Europhys. Lett. {\bf 74}, 574 (2006); R. B. Diener, R. Sensarma, and M. Randeria, Phys. Rev. A {\bf 77}, 023626 (2008).

\bibitem{Stoof} H. T. C. Stoof, K. B. Gubbels, and D. B.M. Dickerscheid, {\em Ultracold Quantum Fields}, (Springer, 2009).

\bibitem{leipralong} L. Jiang, X.-J. Liu, H. Hu, and H. Pu, Phys. Rev. A {\bf 84}, 063618 (2011).	

\bibitem{deviatedipole} J.-Y. Zhang, S.-C. Ji, Z. Chen, L. Zhang, Z.-D. Du, B. Yan, G.-S. Pan, B. Zhao, Y.-J. Deng, H. Zhai, S. Chen, and J.-W. Pan, Phys. Rev. Lett. {\bf 109}, 115301 (2012). 

\bibitem{ram} B. Ramachandhran, B. Opanchuk, X.-J. Liu, H. Pu, P. D. Drummond, and H. Hu, Phys. Rev. A {\bf 85}, 023606 (2012).

\bibitem{wubiao} Q. Zhu, C. Zhang, and B. Wu, Europhys. Lett. {\bf 100}, 50003 (2012).

\bibitem{YuZhai} Z.-Q. Yu, and H. Zhai, Phys. Rev. Lett. {\bf 107}, 195305 (2011).

\bibitem{Iskin} M. Iskin, and A. L. Subasi, Phys. Rev. Lett. {\bf 107}, 050402 (2011).

\bibitem{GongZhang} M. Gong, S. Tewari, and C. Zhang, Phys. Rev. Lett. {\bf 107}, 195303 (2011).

\bibitem{HuiPu} H. Hu, L. Jiang, X.-J. Liu, and H. Pu, Phys. Rev. Lett. {\bf 107}, 195304 (2011). 

\bibitem{hui06rapid} H. Hu, and X.-J. Liu, Phys. Rev. A {\bf 73} 051603(R) (2006).

\bibitem{leocomment} D. Sheehy, and L. Radzihovsky, Phys. Rev. B {\bf 75}, 136501 (2007).

\bibitem{liuwm} R. Liao, Y. Yi-Xiang, and W.M. Liu, Phys. Rev. Lett. {\bf 108}, 080406 (2012).

\bibitem{yi}X. Zhou, G.-C. Guo, W. Zhang, and W. Yi, arXiv:1302.1303 (2013).

\end{thebibliography}
\end{document}